\documentclass[12pt]{article}
\usepackage{epsfig}
\usepackage{subfigure}
\usepackage{amsmath}
\usepackage{amsfonts}
\usepackage{amssymb}
\usepackage{url}
\usepackage{hyperref}
\newcommand{\be}{\begin{equation}}
\newcommand{\ee}{\end{equation}}
\newcommand{\bea}{\begin{eqnarray}}
\newcommand{\eea}{\end{eqnarray}}

\newcommand{\Zp}{{Z^\prime}}
\begin{document}
\begin{titlepage}
\vspace{4\baselineskip}
\begin{center}
{\Large\bf 
Flavorful $Z^\prime$ signatures at LHC and ILC
}
\end{center}
\vspace{1cm}
\begin{center}
{\large 
Shao-Long~Chen~$^{a}$ and Nobuchika~Okada~$^{b,c}$
}
\end{center}
\vspace{0.2cm}
\begin{center}
$^{a}${\it 
Department of Physics and Center for Theoretical Sciences,\\ 
National Taiwan University, Taipei, Taiwan  }\\
$^{b}${\it 
 Department of Physics, University of Maryland, 
 College Park, MD 20742, USA  }\\
$^{c}${\it 
Theory Group, KEK, Tsukuba 305-0801, Japan 
}\\
\medskip
\vskip 10mm
\end{center}
\vskip 10mm
\begin{abstract}
There are lots of new physics models which predict
 an extra neutral gauge boson, referred as $Z^\prime$-boson.
In a certain class of these new physics models,
 the $Z^\prime$-boson has flavor-dependent couplings
 with the fermions in the Standard Model (SM).
Based on a simple model in which couplings of the SM fermions
 in the third generation with the $Z^\prime$-boson
 are different from those of the corresponding fermions
 in the first two generations,
 we study the signatures of $Z^\prime$-boson
 at the Large Hadron Collider (LHC) and
 the International Linear Collider (ILC).
We show that at the LHC, the $\Zp$-boson with mass around 1 TeV
 can be produced through the Drell-Yan processes
 and its dilepton decay modes provide us clean signatures
 not only for the resonant production of $Z^\prime$-boson
 but also for flavor-dependences of the production cross sections.
We also study fermion pair productions at the ILC
 involving the virtual $Z^\prime$-boson exchange.
Even though the center-of-energy of the ILC is much lower
 than a $Z^\prime$-boson mass, the angular distributions and
 the forward-backward asymmetries of fermion pair productions
 show not only sizable deviations from the SM predictions
 but also significant flavor-dependences.
\end{abstract}
\end{titlepage}

\setcounter{page}{1}
\setcounter{footnote}{0}

The search for new physics beyond the Standard Model (SM)
 is one of the most important issues of particle physics today.
In a class of new physics models, the SM gauge group is
 embedded in a larger gauge group and such a model often
 predicts an electrically neutral massive gauge boson,
 referred as $Z^\prime$-boson, associated 
 with the original gauge symmetry breaking into the SM one.
There are many example models such as the left-right symmetric model
 \cite{LR}, Grand Unified Theories based on the gauge groups
 SO(10) \cite{SO10} and $E_6$ \cite{E6}, 
 and string inspired models \cite{string} 
(for a review, see, for example, \cite{ZpRev}).

It will be very interesting if a $Z^\prime$-boson is discovered
at future collider experiments such as the LHC and ILC.
Current limits for the direct production at the Tevatron
 and indirect effects from LEP experiments imply that the $Z^\prime$-boson
 is rather heavy and has a very small mixing with the SM $Z$-boson.
No evidence of a signal has been found,
 and the lower limits on $Z^\prime$ mass at 95\% confidence level
 are set to be in the range from $650$ to $900$ GeV,
 depending on the considered theoretical models~\cite{Abulencia:2006iv}.

Recently studies about measurement of the $Z'$-boson
 at the LHC have been performed \cite{simulations}.
Through the Drell-Yan process, $pp \to  Z^\prime X \to \ell^+\ell^- X$,
 a $Z^\prime$-boson could be discovered at the LHC
 if its mass lies around TeV scale with typical electroweak
 scale couplings to the SM fermions.
Once a $Z^\prime$-boson resonance is observed at the LHC,
 the $Z^\prime$-boson mass can be precisely measured.
The next task is to precisely measure other properties
 of the $Z^\prime$-boson, such as couplings to the SM particles,
 its spin, etc.
Future $e^+e^-$ linear colliders, such as the ILC, will
 be capable of such a task, even if the collider energy
 is not sufficiently high to produce the $Z^\prime$-boson.
For example, the precision goal of the ILC can allow us
 to indicate the existence of $Z^\prime$-boson with mass
 up to 6 times of center-of-mass energies of the collider~\cite{ILC}.

In general, the coupling of $Z^\prime$-boson with the SM
 fermions can be flavor-dependent.
In fact, such a class of models has been proposed
 by many authors~\cite{FCNCZp1, FCNCZp2, Nardi1, Chiang:2007sf}.
If this is the case, the signature of $Z^\prime$-boson should
 show flavor-dependences and the collider phenomenology of
 $Z^\prime$-boson would be more interesting.
In this Letter, we take a simple model recently
 proposed~\cite{Chiang:2007sf},
 where the SM fermions in the third generation have couplings
 with the $Z^\prime$-boson different from those of
 the corresponding fermions in the first two generations,
 and study (flavor-dependent) $Z^\prime$-boson signatures
 at the LHC and ILC.

Let us first give a brief review on a recently proposed
 ``top hypercharge'' model~\cite{Chiang:2007sf}.
This model is based on the gauge group
 SU(3)$_C \times$ SU(2)$_L \times $ U(1)$_{1} \times$ U(1)$_{2}$
 and the SM fermions in the first two generations
 have hypercharges only under U(1)$_1$ while the third generation fermions
 have charges only under U(1)$_2$.
A complex scalar field, $\Sigma$, in the representation
  $({\bf 1}, {\bf 1}, {\bf 1}, 1/2, -1/2)$ is introduced,
 by whose vacuum expectation value (VEV) ($\langle\Sigma\rangle=u/\sqrt{2}$)
 the gauge symmetry U(1)$_{1} \times$ U(1)$_{2}$ is
 broken down to the SM U(1)$_Y$.
Associated  with this gauge symmetry breaking,
 the mass eigenstates of two gauge bosons are described as
\begin{eqnarray}
\left(\begin{array}{c}
B_{\mu}\\
\tilde{B}_{\mu}\end{array}\right)=\left(\begin{array}{cc}
\cos\phi & \sin\phi\\
-\sin\phi & \cos\phi\end{array}\right)\left(\begin{array}{c}
B_{\mu}^{1}\\
B_{\mu}^{2}\end{array}\right)\,,
\end{eqnarray}
where $B_{\mu}^{1.2}$ are the gauge boson of U(1)$_{1,2}$,
 and the mixing angle $\phi$ is defined by
 $\tan\phi=g_{1}^{\prime}/g_{2}^{\prime}$,
 the ratio between the coupling constants of U(1)$_{1,2}$.
The corresponding masses are
 $m_{B_{\mu}}^{2}=0$ and
 $m_{\tilde{B}_{\mu}}^{2}=(g_{1}^{\prime 2}+g_{2}^{\prime 2})u^{2}/4$,
 and the massless state $B_{\mu}$ is nothing
 but the SM U(1)$_Y$ gauge field.

In terms of $B_{\mu}$ and $\tilde{B}_{\mu}$,
 the covariant derivative with respect to
 SU(2)$_L \times$ U(1)$_1 \times$ U(1)$_2$ for a fermion
 with a charge $(Y_1, Y_2)$ under U(1)$_1 \times$ U(1)$_2$
 is given as
\begin{eqnarray}
 D_{\mu}^{f}
 = \partial_{\mu}-igW_{\mu}^{a}T^{a}-ig^{\prime}YB_{\mu}
  -ig^{\prime}(-Y_{1}\tan\phi+Y_{2}\cot\phi)\tilde{B}_{\mu} \, ,
\end{eqnarray}
where $Y=Y_{1}+Y_{2}$ is a hypercharge under the U(1)$_Y$,
 and the U(1)$_Y$ gauge coupling $g^{\prime}$ is defined as
\begin{equation}
 \frac{1}{g^{\prime 2}}
 =\frac{1}{g_{1}^{\prime 2}}+\frac{1}{g_{2}^{\prime 2}}\,.
\end{equation}
We can also express the coupling constants in terms
 of the electron charge $e$ and the analog of
 the weak mixing angle $\theta_{W}$ in the SM as
\begin{equation}
 g=\frac{e}{\sin\theta_{W}}\,,\; g^{\prime}=\frac{e}{\cos\theta_{W}}\,.
\end{equation}

To break the SU(2)$_{L}\times$ U(1)$_{Y}$ gauge symmetry down
 to the U(1)$_{EM}$, two scalar Higgs doublets $\Phi_{1,2}$
 are introduced, which transform under SU(2)$_{L}\times$
 U(1)$_{1}\times$ U(1)$_{2}$ as $({\bf 2},1/2,0)$ and $({\bf 2},0,1/2)$,
 respectively, and develop VEVs as
\begin{equation}
\langle\Phi_{1,2}\rangle=\frac{1}{\sqrt{2}}\left(\begin{array}{c}
0\\
v_{1,2}\end{array}\right)\;.
\end{equation}
Associated with this symmetry breaking, the $W$-boson
 gets mass $M_{W}^{2}=g^{2}(v_{1}^{2}+v_{2}^{2})/4$
 while the mass-squared matrix of the neutral gauge bosons is given by
\begin{equation}\label{mmatrix}
 M_{neutral}^{2}=\frac{u^{2}}{4}
 \left(\begin{array}{ccc}
 0 & 0 & 0\\
 0 & (g^{2}+g^{\prime 2})\epsilon &
 \frac{g^{\prime 2}\epsilon}{\sin\theta_{W}}\cot\phi\cos^{2}\beta\left(\tan^{2}\phi-\tan^{2}\beta\right)\\
0 & ...& \frac{4g^{\prime 2}}{\sin^{2}2\phi}
+g^{\prime 2}\epsilon\left(\cos^{2}\beta\tan^{2}\phi+\sin^{2}\beta\cot^{2}\phi\right)
\end{array}\right)\;,
\end{equation}
 under the basis $(A_{\mu,}Z_{\mu,}\tilde{B}_{\mu})$ with \begin{equation}
\left(\begin{array}{c}
A_{\mu}\\
Z_{\mu}\end{array}\right)=\left(\begin{array}{cc}
\cos\theta_{W} & \sin\theta_{W}\\
-\sin\theta_{W} & \cos\theta_{W}\end{array}\right)\left(\begin{array}{c}
B_{\mu}\\
W_{\mu}^{3}\end{array}\right)\;,
\end{equation}
and
\begin{equation}
\epsilon=\frac{v_{1}^{2}+v_{2}^{2}}{u^{2}}\;\;,\;\;\tan\beta\equiv\frac{v_{2}}{v_{1}}\;.
\end{equation}

There is a mixing between $Z_{\mu}$ and $\tilde{B}_{\mu}$
 which is constrained to be very small $\lesssim$ $10^{-3}$
 by the electroweak precision measurements~\cite{mixing}.
Thus we fix model-parameters to realize such a small mixing, 
 so that the field $\tilde{B}_{\mu}$ is identified as
 the $Z^{\prime}$-boson.
Note that from Eq.~(\ref{mmatrix}), the mixing between
 $Z_{\mu}$ and $\tilde{B}_{\mu}$ vanishes for $\tan^2\phi=\tan^2\beta$.
In the following, we take $\tan^2\phi=\tan^2\beta$, for simplicity.

Assigning $Y_2=0$ for fermions in the first two generations,
 while $Y_1=0$ for fermions in the third generation,
 we obtain the interactions between the $Z^\prime$-boson
 and the SM fermions such as
\begin{equation}
-\mathcal{L}_{Z^{\prime}}=
 \bar{\psi_{f}}\gamma^{\mu}(g_{L}^{f}P_{L}+g_{R}^{f}P_{R})
\psi_{f}Z_{\mu}^{\prime}\; ,
\end{equation}
 where $g_{L,R}$ for each SM fermion are given as
\begin{eqnarray}
g_{L}^{u,d,c,s}=-\frac{1}{6}\frac{e}{\cos\theta_{W}}\tan\phi & ,
& g_{L}^{t,b}=\frac{1}{6}\frac{e}{\cos\theta_{W}}\cot\phi\,; \nonumber \\
g_{L}^{\nu_{e},\nu_{\mu},e,\mu}=\frac{1}{2}\frac{e}{\cos\theta_{W}}\tan\phi & ,
&
g_{L}^{\nu_{\tau},\tau}=-\frac{1}{2}\frac{e}{\cos\theta_{W}}\cot\phi\,;
\nonumber \\
g_{R}^{u,c}=-\frac{2}{3}\frac{e}{\cos\theta_{W}}\tan\phi & ,
& g_{R}^{t}=\frac{2}{3}\frac{e}{\cos\theta_{W}}\cot\phi\,;\\
 \label{couplings}
g_{R}^{d,s}=\frac{1}{3}\frac{e}{\cos\theta_{W}}\tan\phi & ,
& g_{R}^{b}=-\frac{1}{3}\frac{e}{\cos\theta_{W}}\cot\phi\,; \nonumber \\
g_{R}^{e,\mu}=\frac{e}{\cos\theta_{W}}\tan\phi & ,
& g_{R}^{\tau}=-\frac{e}{\cos\theta_{W}}\cot\phi\,; \nonumber \\
 g_{R}^{\nu_{e},\nu_{\mu},\nu_{\tau}}=0 & . \nonumber
\end{eqnarray}
As a result, the $\Zp$-boson couples differently to the first two and the
 third generations.
In general, the family non-universal couplings generate tree-level
 flavor changing neutral currents
 (FCNCs)~\cite{Chiang:2007sf, FCNC1, Lynch:2000md, FCNC2}
 and therefore, they are severely constrained
 by current experimental data.
The constraints on the model parameters due to the FCNC processes
 have been examined in Ref.~\cite{Chiang:2007sf},
 and in our analysis on the $\Zp$-boson phenomenology,
 a parameter set is chosen to be consistent
 with the current experiments.

\begin{figure}[h]
\begin{center}
\includegraphics[width=0.6\textwidth]{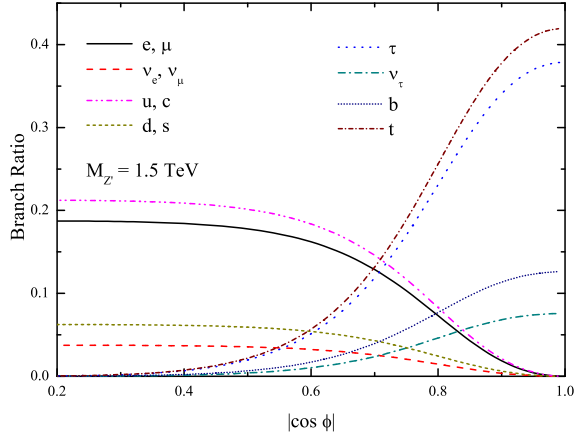}
 \caption{
The branching ratios of the decay $\Zp\to \bar f f$
 as a function of $|\cos \phi|$
 for $M_{\Zp}=1.5$ TeV .
}
\label{fig1}
\end{center}
\end{figure}

The branching ratios of the decay $\Zp \to \bar ff$
 as a function of $|\cos \phi|$ are shown in Fig.~\ref{fig1}.
As can be expected, the branching ratios into the fermions
 in the first two generations are different from those
 into the corresponding fermions in the third generation
 (except for $\tan \phi = 1$).
In the limit, $\cot \phi = 0$, the couplings
 between the $\Zp$-boson and the third generation fermions
 are switched off, while the other limit, $\tan \phi= 0$,
 the couplings with the first two generation fermions vanish.
For $\tan \phi = 1$, the couplings of the third generation fermions
 becomes the same (up to sign) as those of the corresponding
 fermions in the first two generations.
Since the sign difference does not appear in the $\Zp$ decay width, 
 no flavor-dependence can be seen in Fig.~1 for $\tan \phi = 1$.
However, as will be shown below, this sign difference causes
 significant differences at collider phenomenologies.

\begin{figure}[ht]
\begin{center}
\includegraphics[width=0.80 \textwidth]{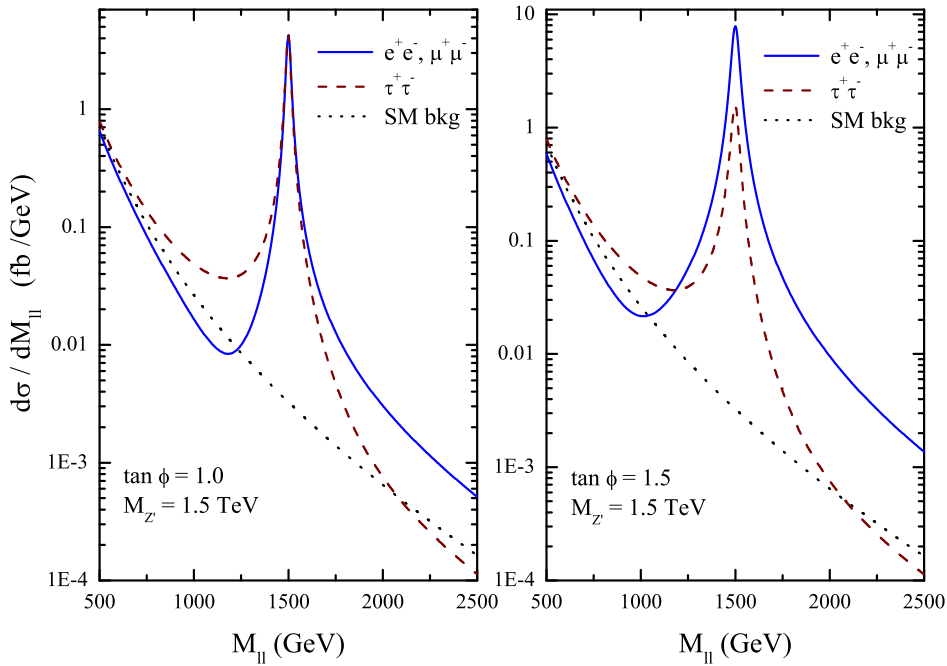}
 \caption{
 The differential cross section for $pp\to \ell^+\ell^-X$
 at the LHC for $M_\Zp=1.5$ TeV and $\tan\phi=1.0$ (left)
 and $1.5$ (right), together with the SM cross section mediated
 by the $Z$-boson and photon (dotted line).
}
\label{fig2}
\end{center}
\end{figure}

Now we investigate the $\Zp$-boson production at the LHC. 
We calculate the dilepton production cross sections
 through the $\Zp$ exchange together with the SM processes
 mediated by the $Z$-boson and photon%
\footnote{
The quark pair production channel, in particular,
 top-quark pair production via the $\Zp$-boson exchange
 is also worth investigating~\cite{Arai},
 since top quark, which electroweakly decays before hadronization,
 can be used as an ideal tool to probe
 new physics beyond the Standard Model~\cite{TopPhys}.
}.
The significance of the $\Zp$-boson discovery
 through the process $pp\to \ell^+\ell^-X$ 
 ($\ell^+\ell^-= e^+ e^-, \mu^+ \mu^-$) 
 has been investigated in Ref.~\cite{Chiang:2007sf}. 
Here we show the dependence of the cross section 
 on the final state invariant mass $M_{ll}$ described as%
\footnote{
For explicit formulas for the production cross section etc,
 see, for example, Appendix in \cite{Arai}.
}
\begin{eqnarray}
 \frac{d \sigma (pp \to \ell^+ \ell^- X) }
 {d M_{ll}}
 &=&  \sum_{a, b}
 \int^1_{-1} d \cos \theta
 \int^1_ \frac{M_{ll}^2}{E_{\rm CMS}^2} dx_1
 \frac{2 M_{ll}}{x_1 E_{\rm CMS}^2}   \nonumber \\
&\times & 
 f_a(x_1, Q^2)
  f_b \left( \frac{M_{ll}^2}{x_1 E_{\rm CMS}^2}, Q^2
 \right)  \frac{d \sigma(\bar{q} q \to \ell^+ \ell^-) }
 {d \cos \theta},
\label{CrossLHC}
\end{eqnarray}
where $E_{\rm CMS} =14$ TeV is the center-of-mass energy of the LHC.
In our numerical analysis, we employ CTEQ5M~\cite{CTEQ}
 for the parton distribution functions
 with the factorization scale $Q= M_{\Zp}$.

\begin{figure}[ht]
\begin{center}
\includegraphics[width=0.8 \textwidth]{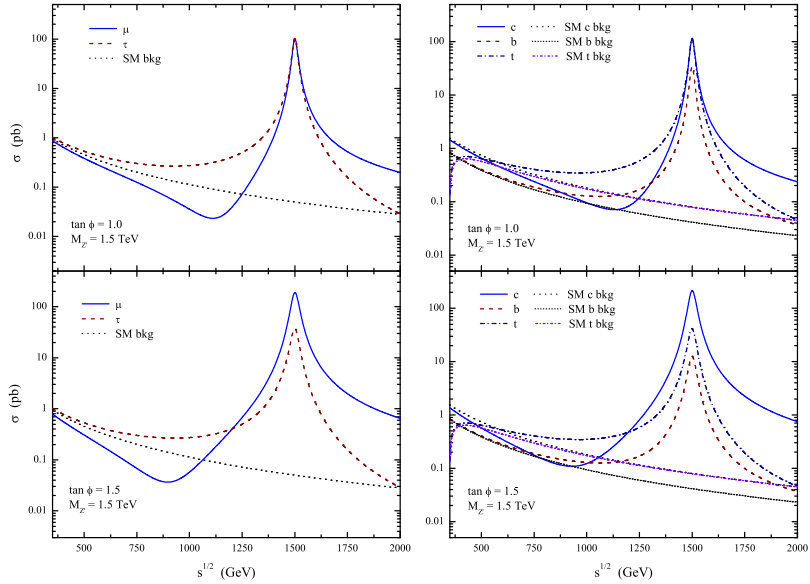}
  \caption{
The cross sections for $e^+ e^- \to {\bar f} f$  at the ILC
 as a function of the center-of-mass energy $\sqrt{s}$.
}\label{fig3}
\end{center}
\end{figure}

Fig.~\ref{fig2} shows the differential cross section
 for $pp \to e^+e^- (\mu^+\mu^-)$ and $\tau^+\tau^-$
 for $M_\Zp=1.5$ TeV and $\tan\phi=1.0$ (left) and $1.5$ (right),
 together with the SM cross section mediated by the $Z$-boson and photon.
Although for $\tan \phi=1$ the peak cross sections are the same,
 the dependence of dilepton invariant mass shows a remarkable
 flavor-dependence.
This is because for the cross sections away from the peak,
 the interference between the $\Zp$-boson mediated process
 and the SM processes dominates and the sign difference
 of the couplings in Eq.~(\ref{couplings})
 between the first two and third generation fermions
 causes the strong flavor-dependence of the cross sections.
For $\tan \phi =1.5$, the flavor-dependence appears
 even in the peak cross sections.

\begin{figure}[ht]
\begin{center}
\includegraphics[width=0.8\textwidth]{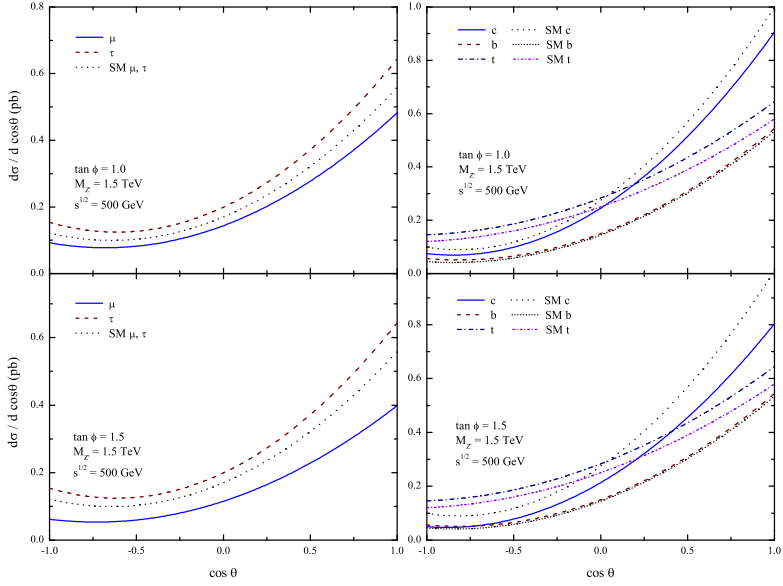}
  \caption{
The differential cross section $d\sigma/d\cos\theta$
 for the process $e^+ e^- \to {\bar f} f$
 at ILC with $\sqrt{s}=500$ GeV.
}\label{fig4}
\end{center}
\end{figure}

When we choose a kinematical region for the invariant mass
 in the range,
 $ M_\Zp-\Gamma_\Zp=1.35$ TeV $\leq M_{ll} \leq M_\Zp=1.5$ TeV,
 for example, $7.8\times 10^3$ and $8.8\times 10^3$ signal events
 would be observed for $e^+e^-$ ($\mu^+ \mu^-$) and $\tau^+\tau^-$ channels,
 respectively, with an integrated luminosity of 100 fb$^{-1}$,
 while the number of evens for the SM background would be about 100. 
In the case $\tan \phi =1.5$, we would expect
 $4.7\times 10^4$ and $9.4\times 10^3$ signal events
 for $e^+e^-$ ($\mu^+ \mu^-$) and $\tau^+\tau^-$ channels,
 respectively, for the kinematical range around the peak,
 $M_\Zp-\Gamma_\Zp=1.35$ TeV $\leq M_{ll} \leq$
 $M_\Zp+\Gamma_\Zp=1.65$ TeV,
 with an integrated luminosity of 100 fb$^{-1}$.

Once a resonance of the $\Zp$-boson has been discovered
 at the LHC, the $\Zp$-boson mass can be determined
 from the peak energy of the dilepton invariant mass.
The difference between the cross sections of different dilepton
 channels at the LHC could be a good distinction between
 flavor-dependent $\Zp$ models and the flavor-universal ones.
The ILC can provide more  precise measurement of the $\Zp$-boson
 properties such as couplings with each (chiral) SM fermion, 
 spin and etc., even if its center-of-mass energy
 is far below the $\Zp$-boson mass~\cite{ILC}.
Then, we next study ILC phenomenologies on $\Zp$-boson.

\begin{figure}[htb!]
\begin{center}
\includegraphics[width=0.6\textwidth]{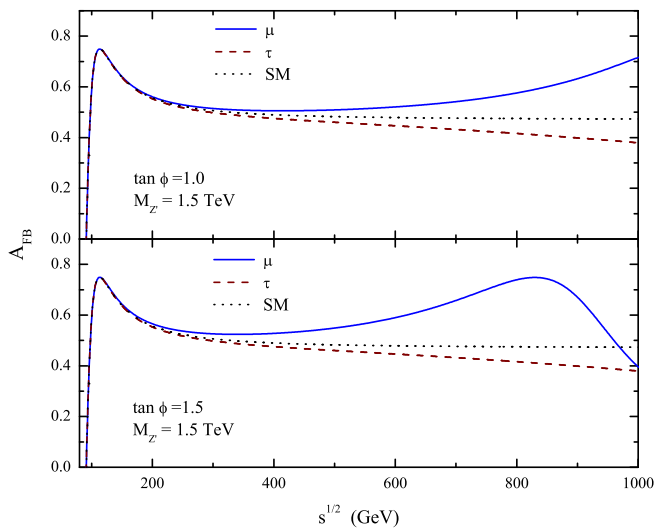}
 \caption{ 
 The forward-backward asymmetry $A^\ell_{FB}$ 
 as a function of the center-of-mass energy $\sqrt{s}$. 
}\label{fig5}
\end{center}
\end{figure}

We begin with calculating the cross sections of
 the process $e^+ e^- \to \bar{f} f$ at the ILC
 with different energies and the results are depicted
 in Fig.~\ref{fig3} for $M_\Zp=1.5$ TeV 
 and $\tan \phi=1$ (upper figures) and $1.5$ (lower figures),
 together with the SM cross sections.
The results show a similar behavior as in Fig.~\ref{fig2}
 and we can see sizable deviations from the SM results
 and also clear flavor-dependences of the cross sections,
 even for $s \ll M_\Zp^2$.
For $\sqrt{s} =500$ GeV fixed, we show the differential cross
 sections for the process $e^+ e^- \to \bar{f} f$ for
 $\tan \phi=1$ (upper figures) and $1.5$ (lower figures) in Fig.~\ref{fig4}.
Since the collider energy is lower than $M_\Zp$, the interference
 between the SM processes and the virtual $\Zp$-boson exchange
 causes the deviations of the cross sections from the SM results.
We can see that the deviations from the SM results are flavor-dependent.
With high integrated luminosity, the deviations could be used
 to determine the $\Zp$ couplings and therefore identify
 the $\Zp$-boson properties.

Furthermore, the forward-backward asymmetries $A^\ell_{FB}$
 are very distinct for different generation dileptons%
\footnote{
 Similar analysis on the model proposed in \cite{Nardi1} 
 was performed in \cite{Nardi2}. 
}
as shown in Fig.~\ref{fig5}. 
Here we have defined the forward-backward asymmetry
 for $e^+e^- \to \ell^+\ell^-$ ($\ell=\mu (e)$ or $\tau$) as
\begin{equation}
 A_{FB}\equiv
 \frac{\int^1_0 \frac{d\sigma}{d\cos\theta}
 d\cos\theta-\int^0_{-1} \frac{d\sigma}{d\cos\theta} d\cos\theta}
 {\int^1_0 \frac{d\sigma}{d\cos\theta}
 d\cos\theta+\int^0_{-1} \frac{d\sigma}{d\cos\theta} d\cos\theta} 
\end{equation}
with $\theta$ being the scattering angle between
 the final state $\ell^-$ and the initial $e^-$ beam directions.
The precise measurements of $A^\ell_{FB}$ at the ILC
 would reveal not only the existence of the $\Zp$-boson
 at very high energy but also its flavor-dependent couplings.

In summary, we have studied the signatures
 of a new charge neutral gauge boson, $\Zp$, at the LHC and ILC.
Such a gauge boson has been predicted in many new physics models
 beyond the SM. In particular, we have concentrated
 on a class of these models where $\Zp$-boson has flavor-dependent
 couplings with the SM fermions.
As a concrete example of such models, we have taken
 the top hypercharge model proposed in~\cite{Chiang:2007sf},
 where the SM fermions in the third generation have couplings
 with the $\Zp$-boson different from those of the corresponding
 SM fermions in the first two generations.
For a $\Zp$-boson mass around 1 TeV,
 the dilepton production cross sections via the $\Zp$-boson
 in  the $s$-channel well-exceed the SM background,
 so that the discovery of the $\Zp$-boson resonance would
 be promising in this case.
In addition, the dependence of the cross sections on
 the dilepton invariant mass shows a clear flavor-dependence
 around the resonance peak and therefore, the flavor-non-universality
 of the $\Zp$-boson resonance could be also observed at the LHC.
We have also analyzed the $\Zp$-boson effects at the ILC.
Even if the energy at the ILC is far below a $\Zp$-boson mass,
 the differential cross sections and the forward-backward asymmetries
 for the fermion pair production processes show
 not only sizable deviations from the SM results but also
 significant flavor-dependences, through which the ILC 
 with a high integrated luminosity could precisely 
 measure the flavor-dependent couplings of the $\Zp$-boson
 with the SM fermions in different generations.
Finally, although the analysis in this Letter was based 
 on the model proposed in~\cite{Chiang:2007sf}, 
 our strategy is applicable to general models 
 of $\Zp$-boson with flavor-dependent couplings. 

\vspace*{0.8cm}
\noindent
{\large {\bf Acknowledgments}}

The work of S.-L.C. was supported in part by the NSC, Taiwan, R.O.C.
He would also like to thank Theory Group, KEK
 for supports and hospitality during his visit.
N.O. would like to thank the Maryland Center 
 for Fundamental Physics, and especially Rabindra N. Mohapatra
 for their hospitality and financial support during his stay.
The work of N.O. is supported in part by 
the National Science Foundation Grant No. PHY-0652363, and 
the Grant-in-Aid for Scientific Research from the Ministry 
of Education, Science and Culture of Japan (No.~18740170).




\end{document}